\documentclass[12pt]{article}
\usepackage{graphicx}
\usepackage{cite}
\usepackage{color}
\def\unit{\mbox{\large \boldmath $1$}}

\newcommand{\bm}[1]{\mbox{\boldmath $#1$}}
\newcommand{\fnd}[2]{\frac{\textstyle #1}{\textstyle #2}}

\newcommand{\xrm}[1]{{\textstyle \mbox{\rm #1}}}

\newcommand{\Real}[1]{\Re {\it e}\left(#1 \right)}
\newcommand{\Imag}[1]{\Im {\it m}\left(#1 \right)}

\begin{document}
\title{Why the Pennington-Wilson expansion with real coefficients
is of little use in the analysis of production processes}
\author{
Eef van Beveren$^{1}$ and George Rupp$^{2}$\\ [10pt]
{\small\it $^{1}$Centro de F\'{\i}sica Te\'{o}rica,
Departamento de F\'{\i}sica,}\\ {\small\it Universidade de Coimbra,
P-3004-516 Coimbra, Portugal}\\ {\small\it http://cft.fis.uc.pt/eef}\\ [10pt]
{\small\it $^{2}$Centro de F\'{\i}sica das Interac\c{c}\~{o}es Fundamentais,
Instituto Superior T\'{e}cnico,}\\
{\small\it Technical University of Lisbon, Edif\'{\i}cio Ci\^{e}ncia,
P-1049-001 Lisboa, Portugal}\\ {\small\it george@ist.utl.pt}\\ [10pt]
{\small PACS number(s): 11.80.Gw, 11.55.Ds, 13.75.Lb, 12.39.Pn}
}

\maketitle

\begin{abstract}
We critically analyse and comment on the claims
of M.~R.~Pennington and D.~J.~Wilson \cite{ARXIV07113521}.
Although we generally agree with their obvious algebra,
it is clearly not applicable to our equations. Moreover, we argue that the
corresponding proposal is not useful for production-data analysis. The
advantages of our approach, which involves complex yet purely kinematical
coefficients, are summarised.
\end{abstract}

The analysis of two-body subamplitudes in processes of strong decay
is often carried out under the spectator assumption
\cite{PR88p1163,PR173p1700,PRD1p2192}.
In such cases, one may express
the two-body production subamplitude \bm{P} as a linear
combination of elements of the two-body scattering amplitude $T$
\cite{PRD35p1633,DAP507p404,PTP99p1031,NPA744p127}.
The latter quantities, which contain the full two-body dynamics,
are then supposed to be known, either from experiment
\cite{AIPCP619p112,NPA679p671,PLB585p200,PRD68p036001,PAN68p1554,
EPJC47p45,IJMPA20p482,HEPPH0606266,PRD74p114001,PLB653p1,EPJC52p55},
or from theoretical considerations
\cite{PLB521p15,PLB527p193,PRD67p014012,EPJC30p503,PLB559p49,EPJA24p437,
HEPPH0703256}.

In Ref.~\cite{ARXIV07064119}, assuming quark-pair creation
within a non-relativistic framework, we deduced a relation between a
subamplitude \bm{P}, describing a meson pair emerging from the products of
a strong three-meson decay process, and the corresponding two-meson
scattering amplitude $T$, reading\footnote{
In Appendix~\ref{defZ} we give the precise relation
between the expressions used in Ref.~\cite{ARXIV07064119}
and \bm{Z}.}
\begin{equation}
\bm{P}\; =\;\Real{\bm{Z}}\; +\; i\, T\,\bm{Z}
\;\;\; ,
\label{Production}
\end{equation}
where \bm{Z} consists of complex functions, smooth in the two-body total
invariant mass, which are of a kinematical origin and do not carry information
on the two-body interactions.
So far, our result~(\ref{Production}) is fully in line
with the conclusions of Ref.~\cite{PRD35p1633}, which was also
based on the OZI rule \cite{OZI} and the spectator picture.
In the latter paper,
it was found that the production amplitude can be written as
a linear combination
of the elastic and inelastic two-body scattering amplitudes,
with coefficients that do not carry any singularities,
but are rather supposed to depend smoothly on
the total CM energy of the system.
However, our result~(\ref{Production}) seems to be in conflict
with yet an extra constraint on \bm{Z},
postulated in Ref.~\cite{PRD35p1633} and
later also in Ref.~\cite{DAP507p404},
namely that the production amplitude should be given
by a \em real \em \/linear combination of the elements
of the transition matrix, owing to the unitarity relation
\begin{equation}
\Imag{\bm{P}}\; =\; T^{\ast}\, \bm{P} \; .
\label{ImaTstera}
\end{equation}
The latter property follows straightforwardly from the operator relations
$\bm{P}V=(1+TG)V=V+TGV=T$, the symmetry of $T$, the realness of $V$,
and the unitarity of $1+2iT$, which gives
$\Imag{\bm{P}}V=\Imag{\bm{P}V}=\Imag{T}=T^{\ast}T=T^{\ast}\bm{P}V$.
This leads, for non-singular potentials $V$,
to Eq.~(\ref{ImaTstera}).

Now, in Ref.~\cite{ARXIV07105823} we proved that our
expression~(\ref{Production}) also satisfies relation~(\ref{ImaTstera}).
However, in Ref.~\cite{ARXIV07113521} M.~R.~Pennington and D.~J.~Wilson
showed how from
Eq.~(\ref{Production}), with the definition
\begin{equation}
\bm{Q}\; =\; T^{-1}\;\Real{\bm{Z}}\; +\; i\,\bm{Z}
\;\;\; ,
\label{RealQ}
\end{equation}
one obtains
\begin{equation}
\bm{P}\; =\; T\,\bm{Q}
\;\;\; .
\label{PeqTQ}
\end{equation}
When one furthermore makes use of the property
\begin{equation}
\Imag{T^{-1}}\; =\; -\,\bm{\unit}
\;\;\; ,
\label{}
\end{equation}
one finds from Eq.~(\ref{RealQ})
\begin{equation}
\Imag{\bm{Q}}\; =\;
\Imag{T^{-1}}\;\Real{\bm{Z}}\; +\;\Real{\bm{Z}}\; =\;
-\,\Real{\bm{Z}}\; +\;\Real{\bm{Z}}\; =\; 0
\;\;\; .
\label{zero}
\end{equation}
Hence, \bm{Q} is a real vector, in agreement with the proof
given in Ref.~\cite{DAP507p404} and the simple demonstration, for
the $2\times2$ case, in Ref.~\cite{ARXIV07113521}.
As a consequence, it appears that \bm{P} can be parametrised
with a real linear combination of $T$-matrix elements (Eq.~\ref{PeqTQ}),
to be contrasted with relation (\ref{Production}), derived in
Ref.~\cite{ARXIV07064119}, where complex coefficients were found.

Before arguing why the result~(\ref{Production}) for relating \bm{P}
and $T$ should be preferred to expression~(\ref{PeqTQ}), we shall first
elaborate on a specific example, namely the coupled $\pi K$+$\eta K$+$\eta'K$
system.

Within the approach of Eq.~(\ref{Production}), we define
\begin{equation}
\left(\begin{array}{c}
P_{\pi K}\\ [10pt] P_{\eta K}\\ [10pt] P_{\eta' K}
\end{array}\right)\; =\;
\left(\begin{array}{l}
\Real{Z_{\pi K}}\, +\,
T_{\pi K\leftrightarrow\pi K}\, iZ_{\pi K}\; +\;
T_{\pi K\leftrightarrow\eta K}\, iZ_{\eta K}\; +\;
T_{\pi K\leftrightarrow\eta' K}\, iZ_{\eta' K}\\ [10pt]
\Real{Z_{\eta K}}\, +\,
T_{\eta K\leftrightarrow\pi K}\, iZ_{\pi K}\; +\;
T_{\eta K\leftrightarrow\eta K}\, iZ_{\eta K}\; +\;
T_{\eta K\leftrightarrow\eta' K}\, iZ_{\eta' K}\\ [10pt]
\Real{Z_{\eta' K}}\, +\,
T_{\eta' K\leftrightarrow\pi K}\, iZ_{\pi K}\; +\;
T_{\eta' K\leftrightarrow\eta K}\, iZ_{\eta K}\; +\;
T_{\eta' K\leftrightarrow\eta' K}\, iZ_{\eta' K}
\end{array}\right)
\;\;\; .
\label{KpiKetaKetap}
\end{equation}
Using Eq.~(\ref{RealQ}) and the definitions in
Eq.~(\ref{KpiKetaKetap}), we obtain, after some straightforward algebra,
for the relation put forward by M.~R.~Pennington and D.~J.~Wilson
(Eq.~(\ref{PeqTQ}))
\begin{eqnarray}
\lefteqn{P_{\pi K}\; =}
\nonumber\\ [10pt] & &
T_{\pi K\leftrightarrow\pi K}\,\left[ iZ_{\pi K}\; +\;
\fnd{1}{\xrm{det}(T)}\,
\left\{ T_{\pi K\leftrightarrow\eta' K}\, T_{\eta' K\leftrightarrow\eta K}
\; -\;
T_{\pi K\leftrightarrow\eta K}\, T_{\eta' K\leftrightarrow\eta' K}\right\}
\,\Real{Z_{\pi K}}\; +
\right.
\nonumber\\ [10pt] & & +\;
\fnd{1}{\xrm{det}(T)}\,
\left\{ T_{\pi K\leftrightarrow\eta' K}\, T_{\eta' K\leftrightarrow\eta K}
\; -\;
T_{\pi K\leftrightarrow\eta K}\, T_{\eta' K\leftrightarrow\eta' K}\right\}
\,\Real{Z_{\eta K}}\; +
\nonumber\\ [10pt] & & \left. +\;
\fnd{1}{\xrm{det}(T)}\,
\left\{ T_{\pi K\leftrightarrow\eta K}\, T_{\eta K\leftrightarrow\eta' K}
\; -\;
T_{\eta K\leftrightarrow\eta K}\, T_{\pi K\leftrightarrow\eta' K}\right\}
\,\Real{Z_{\eta' K}}
\right]\; +
\nonumber\\ [10pt] & & +\;
T_{\pi K\leftrightarrow\eta K}\,\left[ iZ_{\eta K}\; +\;
\fnd{1}{\xrm{det}(T)}\,
\left\{ T_{\eta' K\leftrightarrow\pi K}\, T_{\eta K\leftrightarrow\eta' K}
\; -\;
T_{\eta K\leftrightarrow\pi K}\, T_{\eta' K\leftrightarrow\eta' K}\right\}
\,\Real{Z_{\pi K}}\; +
\right.
\nonumber\\ [10pt] & & +\;
\fnd{1}{\xrm{det}(T)}\,
\left\{ T_{\pi K\leftrightarrow\pi K}\, T_{\eta' K\leftrightarrow\eta' K}
\; -\;
T_{\pi K\leftrightarrow\eta' K}\, T_{\eta' K\leftrightarrow\pi K}\right\}
\,\Real{Z_{\eta K}}\; +
\nonumber\\ [10pt] & & \left. +\;
\fnd{1}{\xrm{det}(T)}\,
\left\{ T_{\pi K\leftrightarrow\eta' K}\, T_{\eta K\leftrightarrow\pi K}
\; -\;
T_{\pi K\leftrightarrow\pi K}\, T_{\eta K\leftrightarrow\eta' K}\right\}
\,\Real{Z_{\eta' K}}
\right]\; +
\nonumber\\ [10pt] & & +\;
T_{\pi K\leftrightarrow\eta' K}\,\left[ iZ_{\eta K}\; +\;
\fnd{1}{\xrm{det}(T)}\,
\left\{ T_{\eta K\leftrightarrow\pi K}\, T_{\eta' K\leftrightarrow\eta K}
\; -\;
T_{\eta' K\leftrightarrow\pi K}\, T_{\eta K\leftrightarrow\eta K}\right\}
\,\Real{Z_{\pi K}}\; +
\right.
\nonumber\\ [10pt] & & +\;
\fnd{1}{\xrm{det}(T)}\,
\left\{ T_{\pi K\leftrightarrow\eta K}\, T_{\eta' K\leftrightarrow\pi K}
\; -\;
T_{\pi K\leftrightarrow\pi K}\, T_{\eta' K\leftrightarrow\eta K}\right\}
\,\Real{Z_{\eta K}}\; +
\nonumber\\ [10pt] & & \left. +\;
\fnd{1}{\xrm{det}(T)}\,
\left\{ T_{\pi K\leftrightarrow\pi K}\, T_{\eta K\leftrightarrow\eta K}
\; -\;
T_{\eta K\leftrightarrow\pi K}\, T_{\pi K\leftrightarrow\eta K}\right\}
\,\Real{Z_{\eta' K}}
\right]
\;\;\; ,
\label{PKpi}
\end{eqnarray}
and similarly for $P_{\eta K}$ and $P_{\eta' K}$.
Hence, by the use of Eq.~(\ref{PeqTQ}), we find
\begin{eqnarray}
Q_{\pi K} & = & iZ_{\pi K}\; +
\fnd{1}{\xrm{det}(T)}\,\left[\;
\left\{ T_{\pi K\leftrightarrow\eta' K}\, T_{\eta' K\leftrightarrow\eta K}
\; -\;
T_{\pi K\leftrightarrow\eta K}\, T_{\eta' K\leftrightarrow\eta' K}\right\}
\,\Real{Z_{\pi K}}\; +
\right.
\nonumber\\ [10pt] & & \hspace{76pt} +\;
\left\{ T_{\pi K\leftrightarrow\eta' K}\, T_{\eta' K\leftrightarrow\eta K}
\; -\;
T_{\pi K\leftrightarrow\eta K}\, T_{\eta' K\leftrightarrow\eta' K}\right\}
\,\Real{Z_{\eta K}}\; +
\nonumber\\ [10pt] & & \hspace{76pt}\left. +\;
\left\{ T_{\pi K\leftrightarrow\eta K}\, T_{\eta K\leftrightarrow\eta' K}
\; -\;
T_{\eta K\leftrightarrow\eta K}\, T_{\pi K\leftrightarrow\eta' K}\right\}
\,\Real{Z_{\eta' K}}
\;\right]
,
\label{QKpi}
\end{eqnarray}
and similarly for $Q_{\eta K}$ and $Q_{\eta' K}$.

An immediate observation, by comparison of Eq.~(\ref{KpiKetaKetap}) with
Eqs.~(\ref{PKpi})(\ref{QKpi}), is that the coefficients $Q_{i}$ in the latter
2 equations contain elements of the $T$ matrix, in contrast with the
entirely kinematical coefficients $Z_{i}$ in Eq.~(\ref{KpiKetaKetap}),
which evidently carry the kinematical scattering cuts.
The separation of the two-body dynamics from its kinematics
is extremely useful for data analysis, whereas coefficients
--- to be fitted in practical production applications ---
that mix dynamics and kinematics do not seem to have
much predictive power in analysing such processes.
Moreover, there is another advantage in using
our result~(\ref{Production}).
Namely, the complex coefficients are not only
just kinematical but even completely known functions
of the two-body CM energy (see Appendix~\ref{defZ}),
leaving any fitting freedom limited
to complex couplings.
The power of this approach has already
been demonstrated \cite{JPG34p1789} in the simple one-channel case,
applied to production processes involving the light scalar mesons
$f_0$(600) (alias $\sigma$) and $K^*_0$(800) (alias $\kappa$),
allowing to dispense with any background contributions.

Finally, there is yet another difficulty with the, in principle,
simple proof of Ref.~\cite{ARXIV07113521}.
In our work on the coupling of confined systems
to scattering channels \cite{AIPCP814p143},
it was shown that the number of degrees of freedom
of the $T$-matrix is equal to the number of confined channels,
which is usually much smaller than
the number of coupled scattering channels.
A very successful comparison with experiment,
based on that observation, was made in Figs.~6 and 7 of
Ref.~\cite{AIPCP814p143}, viz.\ for the coupled $\pi K$+$\eta K$+$\eta' K$
system, showing that there essentially is only one independent eigenphase.
As a consequence, the $T$-matrix is singular.
Hence, relation~(\ref{RealQ}) cannot be applied in such cases.
In our expressions, we carefully avoid the use of $T^{-1}$,
thus ending up with complex coefficients for the relation between
the two-body production subamplitude \bm{P} and the scattering amplitude $T$,
which nevertheless satisfies the unitarity relation~(\ref{ImaTstera}).

Summarising, in Ref.~\cite{ARXIV07064119} we have employed
a microscopic quark-meson model, successful in meson spectroscopy and
non-exotic meson-meson scattering, to derive a simple relation between
production and scattering amplitudes. Although this relation involves an
inhomogeneous real term as well as complex coeffients, all these are purely
kinematical, known functions of the two-body energy. Moreover, the generally
accepted unitarity relation (\ref{ImaTstera}) between production and scattering
is manifestly satisfied.

In Ref.~\cite{ARXIV07113521}, M.~R.~ Pennington and D.~J.~Wilson
criticised our relation by showing, in the $2\times2$ case,
how one can, in principle, rewrite it so as to
end up with the usual real relation, with no inhomogeneous term. In the
foregoing, we have explained why the Pennington-Wilson arguments
are flawed, for two
reasons. First of all, their construction to rewrite our relation for the
$2\times2$ case, which merely amounts to solving two linearly independent real
algebraic equations with two unknowns, involves the inverse of the $T$-matrix.
Now, the latter is generally singular when coupling one or a few $q\bar{q}$
channels to several meson-meson channels, which is compatible with
experimental data on $I\!=\!1/2$ $S$-wave $K\pi$ scattering.

But even if it were possible to define $T^{-1}$,
the Pennington-Wilson construction
would lead to coeffcients that, albeit real, depend on a combination of
$T$-matrix elements, thus mixing dynamics into coefficients usually adjusted
freely in production analyses.
Similar critiques on M.~R.~Pennington's approach
were already formulated by the Ishidas
\cite{AIPCP432p705}.

\section*{Acknowledgements}

This work was supported in part by the {\it Funda\c{c}\~{a}o para a
Ci\^{e}ncia e a Tecnologia} \/of the {\it Minist\'{e}rio da Ci\^{e}ncia,
Tecnologia e Ensino Superior} \/of Portugal, under contract
PDCT/ FP/\-63907/\-2005.

\appendix

\section{Precise definition of \bm{Z_{k}(E)}}
\label{defZ}

In Ref.~\cite{ARXIV07064119} we discussed the partial-wave expansion
of the amplitudes for two-meson production --- together with a spectator
particle --- and scattering, assuming $q\bar{q}$ pair creation.
Hence, the coefficients bear reference to the partial wave $\ell$
and the flavor content  $\alpha$ of the quark pair.
We obtained \cite{ARXIV07064119} the following relation between production and
scattering partial-wave amplitudes:
\begin{equation}
P^{(\ell )}_{\alpha i}\; =\;
g_{\alpha i}\,
j_{\ell}\left( p_{i}r_{0}\right)
\, +\,
i\,\sum_{\nu}\,
g_{\alpha\nu}\,
h^{(1)}_{\ell}\left( p_{\nu}r_{0}\right)\,
T^{(\ell )}_{i\nu}
\;\;\; ,
\label{PAmpdef}
\end{equation}
Accordingly, we must define
\begin{equation}
Z^{(\ell )}_{\alpha k}(E)\; =\;
g_{\alpha k}\,
h^{(1)}_{\ell}\left( p_{k}r_{0}\right)\,
\;\;\; .
\label{Zdef}
\end{equation}
In the latter equations, $j_\ell$ and $h^{(1)}_{\ell}$ stand for the
spherical Bessel function and  Hankel function of the first kind, respectively.
These are smooth functions of the total CM energy, just like
$\mu_{k}$ and $p_{k}$, which are the reduced mass and relative linear
momentum of the two-meson system in the $k$-th channel, respectively.
The constants $g_{\alpha k}$ stand for
the intensities of the $q\bar{q}\to MM$ couplings.
A distance scale $\sim\!0.6$ fm (for light quarks) is represented by $r_{0}$.
In the text we have stripped $Z$ of a reference
to $\ell$ and $\alpha$.

Note, moreover, as can be easily seen
from expressions (\ref{Production}) and (\ref{PAmpdef}),
that the pole structures of the production and scattering
amplitudes are identical, since $\Real{Z_{k}}$, which is proportional to
the spherical Bessel function in Eq.~(\ref{PAmpdef}),
is a smooth function of the total invariant mass.

\newcommand{\pubprt}[4]{{#1 {\bf #2}, #3 (#4)}}
\newcommand{\ertbid}[4]{[Erratum-ibid.~{#1 {\bf #2}, #3 (#4)}]}
\def\AIPCP{AIP Conf.\ Proc.}
\def\DAP{Annalen Phys.}
\def\EPJA{Eur.\ Phys.\ J.\ A}
\def\EPJC{Eur.\ Phys.\ J.\ C}
\def\IJMPA{Int.\ J.\ Mod.\ Phys.\ A}
\def\JPG{J.\ Phys.\ G}
\def\NPA{Nucl.\ Phys.\ A}
\def\PAN{Phys.\ Atom.\ Nucl.}
\def\PLB{Phys.\ Lett.\ B}
\def\PR{Phys.\ Rev.}
\def\PRD{Phys.\ Rev.\ D}
\def\PTP{Prog.\ Theor.\ Phys.}


\begin{thebibliography}{32}
\bibitem{ARXIV07113521}
M.~R.~Pennington and D.~J.~Wilson,
{\it How adding zero to the complex relation between production and scattering
amplitudes found by van Beveren and Rupp gives the expected real relation},
arXiv:0711.3521 [hep-ph].

\bibitem{PR88p1163}
K.~M.~Watson,
{\it  The effect of final state interactions on reaction cross-sections},
\pubprt{\PR}{88}{1163}{1952}.

\bibitem{PR173p1700}
I.~J.~R.~Aitchison and C.~Kacser,
{\it Watson's theorem when there are three strongly interacting particles
in the final state},
\pubprt{\PR}{173}{1700}{1968}.

\bibitem{PRD1p2192}
T.~P.~Coleman, R.~C.~Stafford and K.~E.~Lassila,
{\it Production dependence of the $A_{2}$(1300) mass distribution},
\pubprt{\PRD}{1}{2192}{1970}.

\bibitem{PRD35p1633}
K.~L.~Au, D.~Morgan and M.~R.~Pennington,
{\it Meson dynamics beyond the quark model:
A study of final-state interactions},
\pubprt{\PRD}{35}{1633}{1987}.

\bibitem{DAP507p404}
S.~U.~Chung, J.~Brose, R.~Hackmann, E.~Klempt, S.~Spanier
and C.~Strassburger,
{\it Partial wave analysis in K matrix formalism},
\pubprt{\DAP}{507}{404}{1995}.

\bibitem{PTP99p1031}
M.~Ishida, S.~Ishida and T.~Ishida,
{\it Relation between scattering and production amplitudes:
Concerning $\sigma$ particle in $\pi\pi$ system},
\pubprt{\PTP}{99}{1031}{1998}
[arXiv:hep-ph/9805319].

\bibitem{NPA744p127}
L.~Roca, J.~E.~Palomar, E.~Oset and H.~C.~Chiang,
{\it Unitary chiral dynamics in $J/\psi\to V P P$ decays
and the role of  scalar mesons},
\pubprt{\NPA}{744}{127}{2004}
[arXiv:hep-ph/0405228].

\bibitem{AIPCP619p112}
N.~N.~Achasov,
{\it  Analysis of nature of $\phi\to\gamma\pi\eta$ and
$\phi\to\gamma\pi^{0}\pi^{0}$ decays},
\pubprt{\AIPCP}{619}{112}{2002}
[arXiv:hep-ph/0110059].

\bibitem{NPA679p671}
U.~G.~Mei{\ss}ner and J.~A.~Oller,
{\it  $J/\psi\to\phi\pi\pi$ ($K$ anti-$K$) decays, chiral dynamics
and OZI violation},
\pubprt{\NPA}{679}{671}{2001}
[arXiv:hep-ph/0005253].

\bibitem{PLB585p200}
J.~M.~Link {\it et al.}  [FOCUS Collaboration],
{\it Dalitz plot analysis of $D_{s}^{+}$ and $D^{+}$ decay to
$\pi^{+}\pi^{-}\pi^{+}$ using the $K$-matrix formalism},
\pubprt{\PLB}{585}{200}{2004}
[arXiv:hep-ex/0312040].

\bibitem{PRD68p036001}
I.~Bediaga and M.~Nielsen,
{\it $D_{s}$ decays into $\phi$ and $f_{0}$(980) mesons},
\pubprt{\PRD}{68}{036001}{2003}
[arXiv:hep-ph/0304193].

\bibitem{PAN68p1554}
A.~V.~Anisovich, V.~V.~Anisovich, V.~N.~Markov, V.~A.~Nikonov
and A.~V.~Sarantsev,
{\it Decay $\phi (1020)\to\gamma f_{0}(980)$:
Analysis in the non-relativistic quark model approach},
\pubprt{\PAN}{68}{1554}{2005}
[Yad.\ Fiz.\  {\bf 68}, 1614 (2005)]
[arXiv:hep-ph/0403123].

\bibitem{EPJC47p45}
D.~V.~Bugg,
{\it Reconciling $\phi$ radiative decays with other data for
$a_{0}$(980), $f_{0}$0(980), $\pi\pi\to KK$ and $\pi\pi\to\eta\eta$},
\pubprt{\EPJC}{47}{45}{2006}
[arXiv:hep-ex/0603023].

\bibitem{IJMPA20p482}
P.~Dini  [FOCUS Collaboration],
{\it Dalitz plot analyses from FOCUS},
\pubprt{\IJMPA}{20}{482}{2005}.

\bibitem{HEPPH0606266}
V.~V.~Anisovich,
{\it Once again about the reaction $\phi (1020)\to\gamma\pi\pi$},
arXiv:hep-ph/0606266.

\bibitem{PRD74p114001}
A.~K.~Giri, B.~Mawlong and R.~Mohanta,
{\it  Probing new physics in $B\to f_ {0}(980) K$ decays},
\pubprt{\PRD}{74}{114001}{2006}
[arXiv:hep-ph/0608088].

\bibitem{PLB653p1}
J.~M.~Link {\it et al.} [FOCUS Collaboration] and M.~R.~Pennington,
{\it Dalitz plot analysis of the $D^{+}\to K^{-}\pi^{+}\pi^{+}$ decay
in the FOCUS experiment},
\pubprt{\PLB}{653}{1}{2007}
[arXiv:0705.2248 [hep-ex]].

\bibitem{EPJC52p55}
D.~V.~Bugg,
{\it A study in depth of $f_{0}$(1370)},
\pubprt{\EPJC}{52}{55}{2007}
[arXiv:0706.1341 [hep-ex]].

\bibitem{PLB521p15}
F.~De Fazio and M.~R.~Pennington,
{\it Probing the structure of $f_{0}(980)$ through radiative $\phi$ decays},
\pubprt{\PLB}{521}{15}{2001}
[arXiv:hep-ph/0104289].

\bibitem{PLB527p193}
T.~M.~Aliev, A.~\"{O}zpineci and M.~Savc{\i},
{\it Radiative $\phi\rightarrow\ f_{0}(980)\gamma$ decay in light cone
QCD sum rules},
\pubprt{\PLB}{527}{193}{2002}
[arXiv:hep-ph/0111102].

\bibitem{PRD67p014012}
C.~H.~Chen,
{\it $B\to f_ {0}(980) K^{\ast}$ decays and final state interactions},
\pubprt{\PRD}{67}{014012}{2003}
[arXiv:hep-ph/0210028].

\bibitem{EPJC30p503}
M.~Boglione and M.~R.~Pennington,
{\it Towards a model independent determination of the $\phi\to f_{0}\gamma$
coupling},
\pubprt{\EPJC}{30}{503}{2003}
[arXiv:hep-ph/0303200].

\bibitem{PLB559p49}
P.~Colangelo and F.~De Fazio,
{\it Coupling $g_{f_{0}K^{+}K^{-}}$ and the structure of $f_{0}$(980)},
\pubprt{\PLB}{559}{49}{2003}
[arXiv:hep-ph/0301267].

\bibitem{EPJA24p437}
Yu.~S.~Kalashnikova, A.~E.~Kudryavtsev, A.~V.~Nefediev,
C.~Hanhart and J.~Haidenbauer,
{\it The radiative decays $\phi\to\gamma a_{0}/f_{0}$
in the molecular model for the scalar mesons},
\pubprt{\EPJA}{24}{437}{2005}
[arXiv:hep-ph/0412340].

\bibitem{HEPPH0703256}
M.~R.~Pennington,
{\it Can experiment distinguish tetraquark scalars, molecules
and $\bar{q}q$ mesons?},
arXiv:hep-ph/0703256.

\bibitem{ARXIV07064119}
E.~van Beveren and G.~Rupp,
{\it Relating multichannel scattering and production amplitudes
in a microscopic OZI-based model},
arXiv:0706.4119.

\bibitem{OZI}
S.~Okubo,
{\it $\Phi$ meson and unitary symmetry model},
\pubprt{Phys.\ Lett.}{5}{165}{1963};\\
G.~Zweig,
{\it An $SU_{3}$ model for strong interaction symmetry and its breaking},
CERN Reports TH-401 and TH-412 (1963);\\
see also
{\it Developments in the Quark Theory of Hadrons}, Vol. 1, 22-101 (1981)
editted by D.~B.~Lichtenberg and S.~P.~Rosen;\\
J.~Iizuka, K.~Okada and O.~Shito,
{\it Systematics and phenomenology of boson mass levels (3)},
\pubprt{Prog.\ Theor.\ Phys.}{35}{1061}{1966}.

\bibitem{ARXIV07105823}
E.~van Beveren and G.~Rupp,
{\it The complex relation between production and scattering amplitudes},
arXiv:0710.5823.

\bibitem{JPG34p1789}
E.~van Beveren and G.~Rupp,
{\it $S$-wave and $P$-wave $\pi\pi$ and $K\pi$ contributions
to three-body decay processes in the Resonance-Spectrum Expansion},
\pubprt{\JPG}{34}{1789}{2007}
[arXiv:hep-ph/0703286].

\bibitem{AIPCP814p143}
E.~van Beveren, F.~Kleefeld and G.~Rupp,
{\it Complex Meson Spectroscopy},
{\it XI-th International Conference on Hadron Spectroscopy},
Centro Brasileiro de Pesquisas Fisicas (CBPF),
Rio de Janeiro, Brazil, August 21st - 26th, 2005,
\pubprt{\AIPCP}{814}{143}{2006}
[arXiv:hep-ph/0510120].

\bibitem{AIPCP432p705}
S.~Ishida,
{\it On existence of the $\sigma$(600):
Its physical implications and related problems},
\pubprt{\AIPCP}{432}{705}{1998}
[arXiv:hep-ph/9712229].
\end{thebibliography}
\end{document}